# A new, enhanced diamond single photon emitter in the near infra-red

Igor Aharonovich[1], Chunyuan Zhou[2], Alastair Stacey[1], Julius Orwa[1], David Simpson[3], Andrew D. Greentree[1], François Treussart[2], Jean Francois Roch[2], and Steven Prawer[1]

1. School of Physics, University of Melbourne, Victoria, 3010, Australia
2. Laboratoire de Photonique Quantique et Moléculaire, UMR CNRS 8537, Ecole Normale Supérieure de Cachan, 94235 Cachan cedex, France
3. Quantum Communication Victoria, Victoria, 3010, Australia

**Abstract**

Individual color centers in diamond are promising for near-term quantum technologies including quantum key distribution and metrology. Here we show fabrication of a new color center which has photophysical properties surpassing those of the two main-stay centers, namely the nitrogen vacancy and NE8 centers. The new center is fabricated using focused ion beam implantation of nickel into isolated chemical vapor deposited diamond micro-crystals. Room temperature photoluminescence studies reveal a narrow emission in the near infrared region centered at 768 nm with a lifetime as short as 2 ns. Its focused ion beam compatibility opens the prospect to fabrication with nanometer resolution and realization of integrated quantum photonic devices. Preliminary investigations suggest that this center arises from an as-yet uncharacterized nickel-silicon complex.

(a) Corresponding author i.aharonovich@pgrad.unimelb.edu.au

The discovery in 1997 that diamond color centers could be imaged and investigated at individual level [1] has been highly disruptive to the field of quantum optics. The most well-studied and utilized defect to date has been the negatively-charged nitrogen-vacancy (NV) color center, which has been used for demonstrations of quantum key distribution [2], single photon interference [3], and is the subject of the first commercialized single-photon source [4]. In addition, the potential of NV centers has been identified in a range of applications, including quantum computers and quantum simulators (see for example the recent review [5]), high-precision magnetometry with their unique spin properties [6], and as flexible sources for entangled photons [7]).

Color centers in diamond are highly suitable for such quantum applications because of their combination of desirable properties including photostability, high Debye temperature and room temperature operation. At present, there are three leading single photon sources in diamond. These are the NV center, the nickel related NE8 center [8] and the silicon vacancy Si-V center [9]. The later two exhibit very short lifetime of 2-3 ns, however, Si-V has a long life shelving state which reduces its fluorescence intensity [9]. The narrow emission of the NE8 center around 800 nm is a promising fingerprint to use this center for QKD applications. However, previously demonstrated techniques of incorporation of nickel into CVD diamond films by seeding the substrate with a diamond/nickel powder [10] or implanting the nickel into the substrate onto which the diamond crystals were subsequently grown [11], did not allow a controlled formation of the Ni-related center in a specific diamond crystal and in a preferred location.

Recent advances in diamond materials science and fabrication have raised the possibility to use an ion implantation technique to fabricate color centers in diamond [12, 13]. Focused Ion Beam (FIB) implantation was used to demonstrate a pattern with submicron spatially resolved optical centers in a bulk diamond [14]. Implementing this methodology, and utilizing nickel as the ion source, may address some of the challenges towards a spatially controlled fabrication of nickel related single photon centers in diamond. Furthermore, combining a dual Scanning Electron Microscope (SEM)/FIB to create single photon emitters in a desired crystal and in a preferred position will constitute significant progress towards a scalable control and fabrication of multi qubit quantum optic devices (e.g. Q-switched quantum gates [15]) by affording the ability to locate implantation zones within optical nano-structures. It may even give rise to a 'step and repeat' technology [16] which would be of great benefit to large-scale integrated quantum devices.

In this work we demonstrate the formation of nickel related single photon emitters in a controllable manner by a FIB implantation of nickel into individual CVD diamond nano-crystals. The use of a dual Scanning Electron Microscope SEM/FIB allows for the imaging of a specific crystal prior to implantation with a precise accuracy suitable for past processing and scalability. This method allows fabrication of optical centers in a *specific* crystal of choice as well as in a given space, in contrast to previous approaches where the incorporation of the nickel into diamond crystal was not dependent on the crystal selection [10, 11].

The diamond crystals were grown using a microwave CVD technique following a previously reported procedure [17]. Briefly, the substrate (sapphire or silica cover slip) was seeded with nanodiamond powder (4-6 nm, Nanoamor Inc., Houston, TX, USA) without any ultrasonic treatment followed by a diamond growth using a microwave plasma (900W, 150torr). After growth, the substrate was transferred to a dual SEM/FIB (Jeol JSM 5910) to perform the implantation. The origin of the nickel ions is a Ni/Er liquid metal ion source (LMIS) [18] and the selection of a specific ion type and energy is achieved by applying a set of apertures and an electromagnetic filter. The spot size of the beam can be reduced down to ten nanometers allowing implantation into an individual CVD grown diamond crystal. Nickel ions were implanted with an energy of 30 keV. According to SRIM simulations [19], the stopping range of 30 keV nickel ions is around 20 nm beneath the diamond surface. After implantation, the crystals were annealed at 1000 $^0$C in 95%Ar-5%H$_2$ ambient for 1 hour. Time correlation of photoluminescence intensity was performed with a Hanbury Brown and Twiss (HBT) setup consisting of two avalanche photodiodes in single-photon counting regime (APDs) placed on each side of a beam splitter. For this measurement we used a continuous wave (cw) laser diode excitation source at a wavelength of 687 nm [20]. The collected light from the color centers then passed through a dichroic mirror (DM) and optical filters to remove any residual excitation light. A spectrometer (MicroHR, HORIBA Jobin Yvon) was used to record a photoluminescence spectrum from each center.

Figure 1a shows a schematic of the dual SEM/FIB used to implant the nickel into the diamond crystals. The surface was first imaged by the SEM to determine the diamond density and location. After choosing a specific crystal in a preferred location, a nickel FIB beam with a typical 1-5 pA current was used to perform the implantation into the same crystal. The nickel ions were isolated from the Ni/Er source by a set of electromagnetic lenses and apertures located in the FIB column. The current can be adjusted by moving the apertures in relation to the main ion beam, thus blocking part of the beam. Figure 1b shows an SEM image of the grown diamond crystal. Using our seeding method [17], we are able to control the surface density of crystals on the substrate, thus allowing easy imaging by both the SEM/FIB and the detection using a confocal microscope. The diamonds were grown on a sapphire substrate to reduce the incorporation of silicon into the diamond crystal, which is commonly observed when silicon substrates are used. The diamond crystals were grown to the size of about 200-600 nm, which is small enough to eliminate the total internal reflection of the light, thereby enhancing emitted photon collection efficiency.

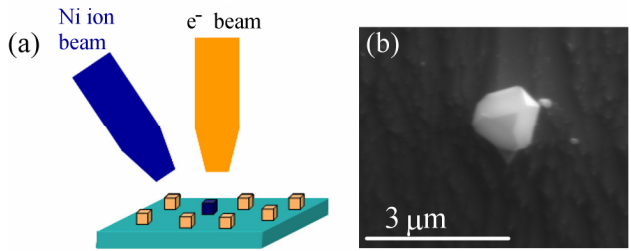

**Figure 1. (color online) (a) Schematic of the SEM/FIB with nickel ion source. The yellow cubes represent diamond crystal while the blue one is the only crystal which was implanted with nickel. (b) SEM image of typical CVD diamond crystal after growth.**

A scan over the post implanted and annealed region using a confocal microscope revealed a bright fluorescence originating from the optical center created within the diamond. Figure 2a shows a photoluminescence (PL) spectrum taken at room temperature from an individual fluorescent diamond crystal, related to the formed nickel color center. A strong narrow band luminescence (FWHM ~ 5 nm) with no significant phonon side band is observed at around 770 nm. More than 90% of the emission is concentrated in the ZPL. Remarkably, this center does not photoluminesce at the same wavelength as the previously reported NE8 or one of its variants [8, 10, 20]. Almost all the implanted crystals revealed a similar emission peak around 770 nm while no PL was detected in the unimplanted regions, convincingly indicating that the origin of the signal is due to the FIB nickel implantation.

The photon statistics of PL light from Ni-related color centers was studied by measuring the normalized second-order time autocorrelation function, $g^{(2)}(\tau)=\langle I(t)I(t+\tau)\rangle/\langle I(t)\rangle^2$, using the HBT interferometer [21]. Figure 2b shows a normalized second order autocorrelation function recorded at room temperature using a cw laser excitation from a color center in an individual diamond crystal located in the implanted area, with the emission spectrum shown in Figure 2a. The dip of $g^{(2)}(\tau)$ at zero delay time indicates that the observed nickel center is indeed a single-photon emitter. Single photon characteristics were found in crystals implanted with a dose of $5\times10^{10}$ Ni/cm$^2$. This dose equals to 10-100 Ni ions per crystal, depending on the exact orientation and the size of the crystal. Higher implantation doses produced double emitters with a $g^{(2)}(0) \sim 0.5$, and emitters exhibiting classical emission, i.e. $g^{(2)}(0) \sim 1$.

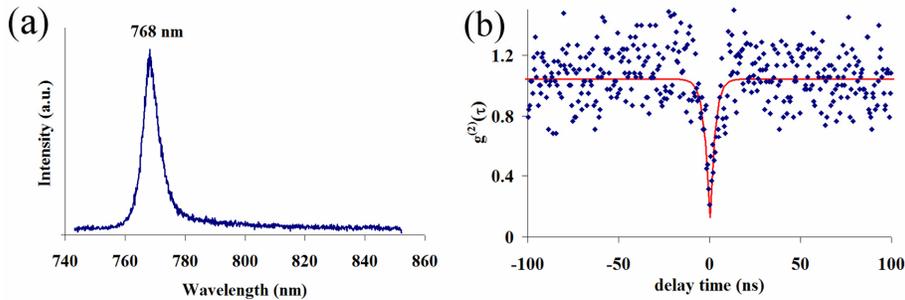

**Figure 2. (a) PL spectrum from individual nickel implanted diamond crystal under 687 nm excitation recorded at room temperature. (b) Corresponding normalized time autocorrelation function $g^{(2)}(\tau)$ of the PL signal recorded at room temperature. The excitation power is 60 μW. The dip at zero delay time with $g^{(2)}(0)=0.2$, indicates single photon emission. The dots indicate the experimental data while the solid line is a fit of $g^{(2)}(\tau)$ taking into account the instrumental response function due to finite time resolution of the APDs and the correlation electronics.**

We also recorded the $g^{(2)}(\tau)$ function with increased cw laser excitation power to measure the decay time of the emitter. By extrapolating a linear fit to a zero excitation power the lifetime of the center is calculated to be as short as 2 ns. This short value is 6 times faster than the typical lifetime of an NV center in a CVD crystal and thus makes the nickel related center attractive for quantum photonic applications with a theoretical emission rate as fast as 0.5 GHz operating at room temperature. This value is very close to currently developed high speed QKD systems with clock rates of 2 GHz [22]. The fluorescence intensity was measured as a function of laser power to evaluate the emission rate at saturation (Figure 3). A high emission rate of 140 kcounts/s was measured with only one APD on the HBT setup. From the fit curve (Figure 3), the total count rate at saturation power is therefore estimated to be greater than 200 kcounts/s. Optimization of the diamond growth substrate, being for example a highly reflective mirror [2] or coupling to waveguide structures [23], should then lead to the realization of a very efficient single photon source

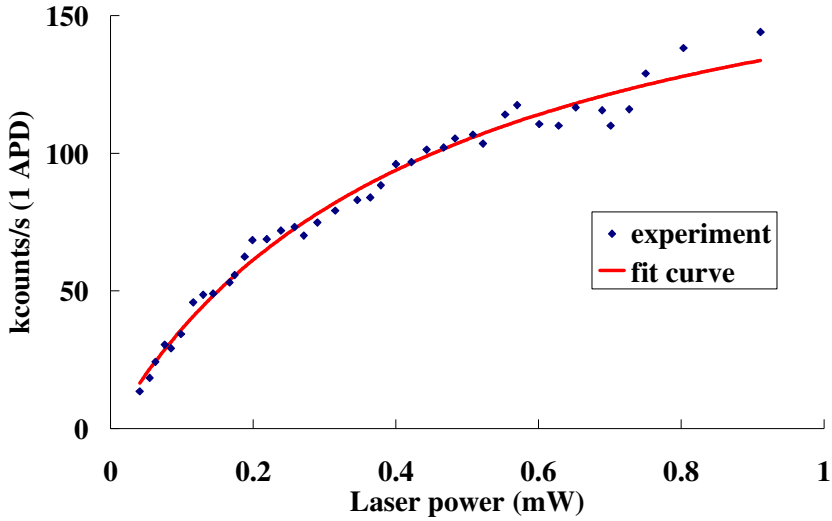

**Figure 3. (Color online) Counting rate data corresponding to the background corrected PL intensity from the same single Ni-related color center shown on figure 2, as a function of excitation power. The background counting rate is 1.8 kcounts/s for an excitation power of 0.9 mW.**

The detection efficiency was then inferred following the procedure described in ref [20] leading to an estimated total detection efficiency of $\eta_{det} \approx 0.04\%$. The roughly estimated photoluminescence quantum yield is therefore $\eta_Q \approx \dfrac{200\,kcps \cdot 2\,ns}{0.04\%} = 100\%$. This latter result strongly suggests that the obtained optical center is a two level system without any associated shelving or a third state with non-radiative decay as observed for NV [2], Si-V [9] and NE8 [10] centers.

Finally, we discuss the atomistic structure of the centers. It is clear from our experimental data that the observed center is not the previously reported nickel nitrogen, NE8 center. The luminescence was observed only from implanted crystals, and therefore it is likely that the centers are nickel related. Previous work reported various emission lines characteristic of nickel impurities in diamond [24], however, an emission around 770 nm has not been observed or assigned to any particular nickel related center so far. Besides nitrogen, the most common and favorably incorporated impurity is silicon, which forms a common Si-V center within the diamond lattice. Therefore, we assume that the new center is related to both nickel and silicon complexes. To check our assumptions, we performed a reference experiment by implanting both silicon and nickel into pure synthetic, type IIa, 3×3×0.5 mm diamond produced by Element 6 (e6 Inc., UK). Nickel ions were implanted with an energy of 37.5 keV while silicon ions were implanted using 25 keV. According to SRIM simulations [19], the stopping range of both nickel and silicon ions is around 18 nm beneath the diamond surface. The implantation was followed by the same annealing procedure as used on our diamond micro-crystals. A similar PL line centered at 766 nm was observed (Figure 4) while a nickel only implantation into the same crystal did not result in the formation of this specific emission line. Note that the doublet at around 883/885 nm is associated with an interstitial nickel defect in diamond [25] and appears in both performed nickel implantations. These observations provide solid evidence that the new emission line at around 770 nm is due to a complex containing both nickel and silicon. Note that those two atoms create a large distortion in the diamond crystal. Hence, it is possible that upon annealing a vacancy is combined with those two atoms and the actual structure consists of Si and Ni impurities associated to a vacancy. Given that the CVD diamonds do not show NV emission, it is unlikely that the complex also contains N. The source of the silicon inside the grown CVD diamond crystals may possibly originate from the quartz made CVD chamber. An incorporation of a few ppm of silicon into the diamond crystals during the growth is hence likely. The concentration might be too low to be observed as Si-V because that center is not an efficient emitter but when it complexes with an implanted nickel ion, it may lead to the formation of the observed very bright centers, even for very low silicon concentrations. Investigations of the atomistic structure and theoretical simulations will be the subject of future work.

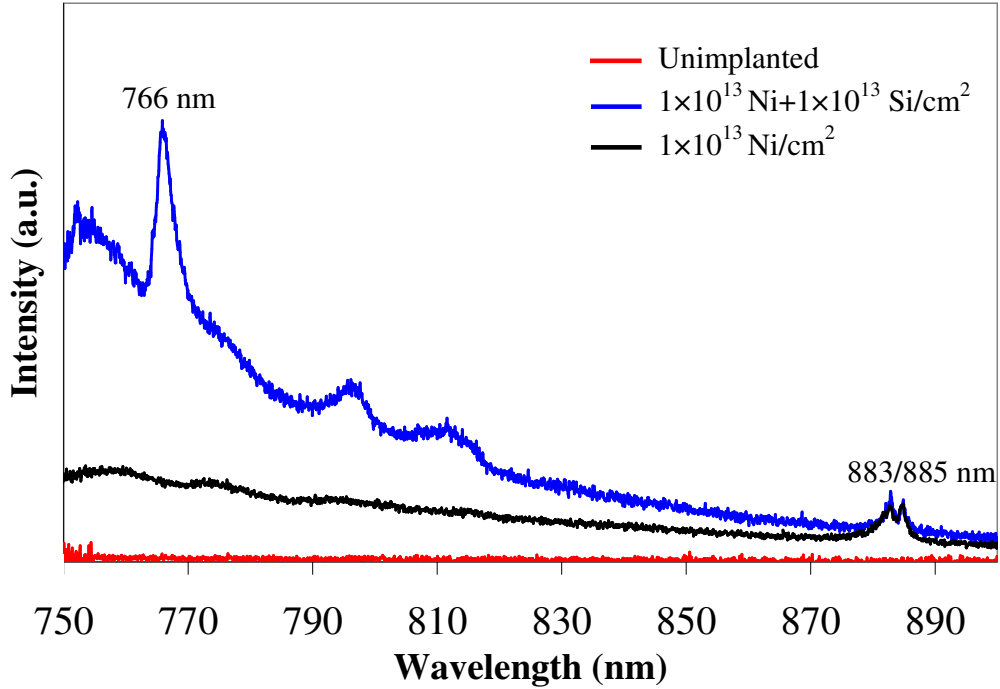

**Figure 4. (Color online) PL spectra from Ni only (black) and a co implantation of Ni/Si (blue) into type IIa e6 CVD diamond. The measurement is taken at 77K under 514 nm excitation. The red curve is recorded from the non implanted region.**

In summary, we have engineered a highly efficient single photon emitter based on nickel and silicon complex in individual CVD diamond crystal. A remarkable lifetime of only 2 ns, a total high count rate (from two APDs) of 280 kHz, and the operation at room temperature makes it a perfect solid state candidate to be used in integrated quantum optics. The implementation of FIB for Ni implantation into selected diamond crystals is an important step towards the fabrication of complex engineered photonic structures. This comprehensive technique along with co-implantation of two or more ions in the same spot, opens new possibilities towards a tailored fabrication of optical building blocks for quantum optical engineering networks. These are major steps towards scalability and design of various optical light guide cavities which must couple individual photons with minimum loss.

**Acknowledgments**

This work was supported by the Australian Research Council, The International Science Linkages Program of the Australian Department of Innovation, Industry, Science and Research and by the European Union under the EQUIND (IST-034368) and NEDQIT (ERANET Nano-Sci) projects, and by the French Agence Nationale de la Recherche PROSPIQ project (ANR-06-NANO-041). ADG is the recipient of an Australian Research Council Queen Elizabeth II Fellowship (project No. DP0880466). I Aharonovich acknowledges ARNAM for their financial support and C Zhou acknowledges support from the Fondation EADS.

**References:**

1. A. Gruber, A, Drabenstedt, C Tietz, L Fleury, J. Wrachtrup, C von Borczyskowski, Science, 276, 2012 (1997).

2. A. Beveratos, R. Brouri, T. Gacoin, A. Villing, J.-P. Poizat, and P. Grangier, Phys. Rev. Lett., 89, 187901, (2002).

3. V. Jacques, E Wu, F. Grosshans, F. Treussart P. Grangier, A. Aspect, and J.-F. Roch, Science, 315, 966 (2007)

4. http://qcvictoria.com

5. A. D. Greentree, B. A. Fairchild, F. M. Hossain and S. Prawer, Materials Today, 11, 22 (2008).

6. C. L. Degen, Appl. Phys. Lett. 92, 243111 (2008); J. R. Maze, P. L. Stanwix, J. S. Hodges, S. Hong, J. M. Taylor, P. Cappellaro, L. Jiang, M. V. Gurudev Dutt, E. Togan, A. S. Zibrov, A. Yacoby, R. L. Walsworth, and M. D. Lukin, Nature 455, 644 (2008); G. Balasubramanian, I. Y. Chan, R. Kolesov, M. Al-Hmoud, J. Tisler,

C. Shin, C. Kim, A. Wojcik, P. R. Hemmer, A. Krueger, T. Hanke, A. Leitenstorfer, R. Bratschitsch, F. Jelezko, J. Wrachtrup, Nature 455, 648 (2008);

7. S. J. Devitt, A. D. Greentree, R. Ionicioiu, J. L. O'Brien, W. J. Munro, and L. C. L. Hollenberg, Phys. Rev. A 76, 052312 (2007).

8. T. Gaebel, I. Popa, A. Gruber, M. Domhan, F. Jelezko, and J. Wrachtrup, New J. Phys. 6, 98 (2004).

9. C. Wang, C. Kurtsiefer, H. Weinfurter and B. Burchard, J. Phys. B: At. Mol. Opt. Phys. 39, 37 (2006).

10. J. Rabeau, Y. Chin, S. Prawer, F. Jelezko, T. Gaebel, and J. Wrachtrup, Appl. Phys. Lett. 86, 131926 (2005).

11. I. Aharonovich, C. Zhou, A. Stacey, F. Treussart, J.-F. Roch and S. Prawer, Appl. Phys. Lett. (2008) in press.

12. F. C. Waldermann, P. Olivero, J. Nunn, K. Surmacz, Z. Y. Wang, D. Jaksch, R. A. Taylor, I. A. Walmsley, M. Draganski, P. Reichart, A. D. Greentree, D.N. Jamieson, S. Prawer, Diam. Rel. Mater. 16, 1887 (2007).

13. J. R. Rabeau, P. Reichart, G. Tamanyan, D. N. Jamieson, S. Prawer, F. Jelezko, T. Gaebel, I. Popa, M. Domhan, and J. Wrachtrup, Appl. Phys. Lett. 88, 023113 (2006).

14. J. Martin, R. Wannemacher, J. Teichert and L. Bischoff and B. Kohler, Appl. Phys. Lett., 75, 3096, (1999).

15. C.-H. Su, A. D. Greentree, W. J. Munro, K. Nemoto, and L. C. L. Hollenberg, Phys. Rev., A (in press), arXiv:0809.2133.

16. D. N. Jamieson, C. Yang, T. Hopf, S. M. Hearne, C. I. Pakes, and S. Prawer M. Mitic, E. Gauja, S. E. Andresen, F. E. Hudson, A. S. Dzurak, and R. G. Clark, Appl. Phys. Lett. **86**, 202101 (2005).

17. A. Stacey, I. Aharonovich, S. Prawer and J. E. Butler, DOI:10.1016/j.diamond.2008.09.020

18. L. Bischoff, Ultramicroscopy 103, 59, (2005).

19. The Stopping and Range of Ions in Matter, www.srim.org

20. E Wu, V. Jacques, H. Zeng, P. Grangier, F. Treussart and J.-F. Roch Opt. Express., 14, 1297 (2006).

21. R. Brouri, A. Beveratos, J.-Ph. Poizat, and P. Grangier, Opt. Lett. 25, 1294–1296 (2000).

22. K. J. Gordon, V. Fernandez, G. S. Buller, I. Rech, S. D. Cova, P. D. Townsend, Opt. Express. 13, 3015 (2005).

23. K.-M.C. Fu, C. Santori, P.E. Barclay, I. Aharonovich, S. Prawer, N. Meyer, A. M. Holm, and R.G. Beausoleil, Appl. Phys. Lett., 93, 234107, (2008).

24. A. Yelisseyev, S. Lawson, I. Sildos, A. Osvet, V. Nadolinny, B. Feigelson, J.M. Baker, M. Newton, O. Yuryeva, Diam. Rel. Mater. 12, 2147 (2003).

25. K. Iakoubovskii and G. Davies Phys. Rev. B, **70**, 245206 (2004) and references within.